\documentclass[12pt]{article}
\usepackage{amssymb}
\usepackage{amsmath}

\begin{document}

\title{ \begin{flushright}
{ \small CECS-PHY-06/11 }
\end{flushright}
\vskip 1.0cm Electrically charged black hole with scalar hair}
\author{Cristi\'{a}n Mart\'{\i}nez and Ricardo Troncoso \\
\\
{\small Centro de Estudios Cient\'{\i}ficos (CECS), Casilla 1469, Valdivia,
Chile.}\\
{\small {e-mails:} \texttt{{\footnotesize {martinez@cecs.cl, ratron@cecs.cl }
}}}}
\date{}
\maketitle

\begin{abstract}
An electrically charged black hole solution with scalar hair in four
dimensions is presented. The self-interacting scalar field is real and it is
minimally coupled to gravity and electromagnetism. The event horizon is a
surface of negative constant curvature and the asymptotic region is locally
an AdS spacetime. The asymptotic fall-off of the fields is slower than the
standard one. The scalar field is regular everywhere except at the origin,
and is supported by the presence of electric charge which is bounded from
above by the AdS radius. In turn, the presence of the real scalar field
smooths the electromagnetic potential everywhere. Regardless the value of
the electric charge, the black hole is massless and has a fixed temperature.
The entropy follows the usual area law. It is shown that there is a
nonvanishing probability for the decay of the hairy black hole into a
charged black hole without scalar field. Furthermore, it is found that an
extremal black hole without scalar field is likely to undergo a spontaneous
dressing up with a nontrivial scalar field, provided the electric charge is
below a critical value.
\end{abstract}

\section{Introduction}

It has been recently shown that General Relativity with a minimally coupled
self-interacting real scalar field, in four dimensions, admits an exact
hairy black hole solution \cite{MTZ-Top}. This result is somehow unexpected
since it circumvents the so called no-hair conjecture, which originally
stated that a black hole should be characterized only in terms of its mass,
angular momentum and electric charge \cite%
{Ruffini-Wheeler,Bekenstein:1971hc,Teitelboim-No-hair} (for recent
discussions see e.g., \cite{review}). The thermodynamical analysis of the
hairy black hole reveals the existence of a second order phase transition,
so that below a critical temperature, a black hole in vacuum undergoes a
spontaneous dressing up with a nontrivial scalar field. According to the
analysis of Ref. \cite{Winstanley-CQG} the solution is expected to be stable
against linear perturbations, and progress towards the proof of nonlinear
stability for solutions with the same asymptotic behavior has been performed
in \cite{Hertog-Hollands}. It can also be seen that below the critical
temperature, these hairy black holes curiously admit only a finite number of
quasinormal modes \cite{QN-Greece}. Furthermore, this black hole can be
uplifted as a solution of eleven-dimensional supergravity \cite{Papa11}. The
black hole solution in \cite{MTZ-Top} can be seen as the neutral case of the
electrically charged hairy black hole found in \cite{MST}.

Numerical hairy black hole solutions of this sort have also been found in
Refs. \cite{Torii:2001pg,Winstanley:2002jt,HM1,Hertog-MaedaSTAB}, and
another exact solution was found in \cite{Zlo}. In the case of conformally
coupled scalar fields, exact black hole solutions were known to exist since
the 70's \cite{BBMB}, however the scalar field in this case diverges at the
horizon. This last obstacle can be avoided considering a cosmological
constant and a quartic self-interaction term for the scalar field \cite%
{dSBH,MTZ-Top,MST}. The thermodynamics and the stability for the black hole
with positive cosmological constant have been discussed in Refs. \cite%
{ThermodSBH} and \cite{Instability}, respectively. Further aspects of this
class of solutions have been studied in \cite{Further-aspects}, and
numerical black hole solutions for nonminimally coupled scalar fields have
been found in Refs. \cite{Winstanley:2002jt,Winstanley-CQG,Winstanley+Radu}.
In three dimensions, black holes dressed with conformally and minimally
coupled scalar fields were found in \cite{Martinez:1996gn,HMTZ-2+1}, and
some of their properties were analyzed in Refs. \cite{GMT,Further3}.

In this paper, an electrically charged black hole solution of gravity
minimally coupled to a real self-interacting scalar field and
electromagnetism in four dimensions is presented. The self interacting
potential induces a negative cosmological constant which allows the event
horizon to be a surface of negative constant curvature which surrounds the
singularity at the origin, and the asymptotic region is locally an AdS
spacetime. The self-interacting potential considered here differs from the
one considered in \cite{MTZ-Top}, but it has the same mass term\footnote{%
The comparison can be precisely seen in the conformal frame. In this case
the action is mapped to gravity with a negative cosmological constant and a
conformally coupled scalar field without selfinteraction, while in Ref.\cite%
{MTZ-Top}, a quartic selfinteraction term with a fixed coupling constant was
considered.}.

The effect of having asymptotically AdS black holes solutions whose horizon
has a nontrivial topology is known to occur in vacuum \cite{Lemos,
Vanzo:1997gw}, as well as in the presence of the electromagnetic field \cite%
{Cai-Zhang,Brill-Louko-Peldan}. For the black hole solution presented here,
the asymptotic fall-off of the fields at the asymptotic region is slower
than the standard one, as in \cite{Henneaux-Teitelboim}, for a localized
distribution of matter. The scalar field is regular everywhere except at the
origin, and is supported by the presence of electric charge which is bounded
from above by the AdS radius. In turn, the presence of the real scalar field
smooths the electromagnetic potential everywhere. The thermodynamics is
discussed in section \ref{thermo}, where it is found that regardless the
value of the electric charge, the black hole is massless and it has a fixed
temperature. The entropy follows the usual area law, as expected. In section %
\ref{thermal}, it is shown that there is a nonvanishing probability for the
decay of the hairy black hole into a black hole without scalar field.
Possible decays including the extremal bare black hole are also analyzed,
and it is found that an extremal black hole without scalar field is likely
to undergo a spontaneous dressing up with a nontrivial scalar field,
provided the electric charge is below certain critical value. Section \ref%
{concluding} is devoted to some concluding remarks.

\section{Black hole solution}

\label{thesolution}

Let us consider gravity minimally coupled to a real self-interacting scalar
field and electromagnetism in four dimensions. The action is given by 
\begin{equation}
I[g_{\mu \nu },\phi ,A_{\mu }]=\int d^{4}x\sqrt{-g}\left[ \frac{1}{16\pi G}R-%
\frac{1}{2}g^{\mu \nu }\partial _{\mu }\phi \partial _{\nu }\phi -V(\phi )-%
\frac{1}{16\pi }F^{\mu v}F_{\mu \nu }\right] \;,  \label{action}
\end{equation}%
where $G$ is the Newton constant, and the self-interaction potential is
chosen as 
\begin{equation}
V(\phi )=-\frac{3}{8\pi Gl^{2}}\cosh ^{4}\left( \sqrt{\frac{4\pi G}{3}}\phi
\right) \;.  \label{Potential}
\end{equation}%
This potential has a global maximum at $\phi =0$, giving rise to a negative
cosmological constant, which can be written in terms of the AdS radius as $%
\Lambda =-3l^{-2}$. Its mass term, $m^{2}=\left. V^{\prime \prime
}\right\vert _{\phi =0}=-2l^{-2}$, satisfies the Breitenlohner-Freedman
bound, which ensures the perturbative stability of AdS spacetime \cite{B-F}.

The field equations are given by 
\begin{eqnarray}
G_{\mu \nu } &=&8\pi G\,\left( T_{\mu \nu }^{\phi }+T_{\mu \nu }^{em}\right)
\;,  \label{Einstein-eq} \\
\square \phi  &=&\frac{dV}{d\phi }\;,  \label{Scalar-eq} \\
\partial _{\nu }\left( \sqrt{-g}F^{\mu \nu }\right)  &=&0\ ,
\label{Maxwell-eq}
\end{eqnarray}%
where the scalar and electromagnetic pieces of stress-energy tensor are 
\begin{eqnarray}
T_{\mu \nu }^{\phi } &=&\partial _{\mu }\phi \partial _{\nu }\phi -\frac{1}{2%
}g_{\mu \nu }g^{\alpha \beta }\partial _{\alpha }\phi \partial _{\beta }\phi
-g_{\mu \nu }V(\phi ),  \label{Tuv-Phi} \\
T_{em}^{\mu \nu } &=&-\frac{1}{4\pi }\left( F_{\ \alpha }^{\mu }F^{\alpha
\nu }+\frac{1}{4}g^{\mu \nu }F^{\alpha \beta }F_{\alpha \beta }\right) ,
\label{Tuv-em}
\end{eqnarray}%
respectively.

The field equations are solved by the following static metric: 
\begin{eqnarray}
ds^{2} &=&-\left( 1+\frac{Gq^{2}}{r^{2}}\right) ^{-1}\left( \frac{r^{2}}{%
l^{2}}-1+\frac{Gq^{2}}{l^{2}}\right) dt^{2}  \notag \\
&+&\left( 1+\frac{Gq^{2}}{r^{2}}\right) ^{-2}\left( \frac{r^{2}}{l^{2}}-1+%
\frac{Gq^{2}}{l^{2}}\right) ^{-1}dr^{2}+r^{2}d\sigma ^{2}\;,
\label{Black-Hole}
\end{eqnarray}%
provided the scalar field and the electromagnetic potential are given by 
\begin{eqnarray}
\phi  &=&\sqrt{\frac{3}{4\pi G}}\;\mbox{Arctanh}\sqrt{\frac{Gq^{2}}{%
r^{2}+Gq^{2}}}\;,  \label{scalar} \\
A &=&-\frac{q}{\sqrt{r^{2}+Gq^{2}}}dt\ ,  \label{A}
\end{eqnarray}%
respectively. In (\ref{Black-Hole}), $d\sigma ^{2}$ is the line element of
the base manifold $\Sigma $, which has negative constant curvature (rescaled
to $-1$), so that is locally isometric to the hyperbolic manifold $H^{2}$.
Thus, a smooth base manifold $\Sigma $ can be obtained through a quotient of
the form $\Sigma =H^{2}/\Gamma $, where $\Gamma $ is a freely acting
discrete subgroup of $O(2,1)$. The metric (\ref{Black-Hole}) describes an
asymptotically locally AdS spacetimes, and if $\Sigma $ is assumed to be
compact without boundary, it has a single timelike Killing vector given by $%
\partial _{t}$.

The integration constant $q$, corresponds to the electric charge which is
given by 
\begin{equation}
Q=\frac{\sigma }{4\pi }q\;,  \label{BlackHoleMass}
\end{equation}%
where $\sigma $ denotes the area of $\Sigma $, and as it is shown below, the
mass of this solution vanishes for any value of $q$.

The curvature and the scalar field are singular at the origin $r=0$, but the
electromagnetic potential is regular everywhere.

The metric (\ref{Black-Hole}) describes a black hole solution with topology $%
\mathbb{R}^{2}\times \Sigma $, with an event horizon located at 
\begin{equation}
r_{+}=\sqrt{l^{2}-Gq^{2}}\ ,  \label{horizon}
\end{equation}%
provided the electric charge is bounded from above by 
\begin{equation}
q^{2}<\frac{l^{2}}{G}\;.  \label{Mass-range}
\end{equation}%
For $q^{2}=l^{2}/G$, the spacetime has a nut on the null curve $r=0$, which
coincides with the singularity.

This black hole has the same causal structure as the Schwarzschild-AdS black
hole, where at each point of the Penrose diagram the sphere is replaced by $%
\Sigma $. The horizon radius satisfies the bound $r_{+}\leq l$, which is
saturated for $q=0$. Note that the scalar field cannot be switched off
keeping the electric charge fixed. This means that the scalar field can be
switched off only if the electric charge vanishes, and then the metric
reduces to%
\begin{equation}
d\bar{s}^{2}=-\left( \frac{r^{2}}{l^{2}}-1\right) dt^{2}+\left( \frac{r^{2}}{%
l^{2}}-1\right) ^{-1}dr^{2}+r^{2}d\sigma ^{2}\ ,  \label{muzero}
\end{equation}%
which corresponds to a negative constant curvature spacetime\footnote{%
Spacetimes of the form (\ref{muzero}) admit Killing spinors provided $\Sigma 
$ is a noncompact surface \cite{Aros:2002rk}. In this case, the metric
describes the supersymmetric ground state of a warped black string. Its
stability under gravitational perturbations has been explicitly proved in 
\cite{Gibbons:2002pq}.}.

The electromagnetic potential at the origin has a fixed value that only
depends on the sign of the electric charge and the Newton constant%
\begin{equation*}
\left. A_{t}\right\vert _{r=0}=-\frac{\mathrm{sgn}(q)}{\sqrt{G}}\ ,
\end{equation*}%
and at the horizon is given by $\left. A_{t}\right\vert _{r=r_{+}}=-q/l$. It
is remarkable that the presence of the real scalar field produces a back
reaction on the metric that regularize the Maxwell field everywhere. The
field strength two-form reads%
\begin{equation*}
F=q\frac{r}{\left( r^{2}+Gq^{2}\right) ^{3/2}}dr\wedge dt\ ,
\end{equation*}%
which at the origin grows linearly as $\mathrm{sgn}(q)q^{-2}G^{-3/2}\ r$,
has an extremum at $r=|q|\sqrt{G/2}$ given by $2(3\sqrt{3}Gq)^{-1}$, and
asymptotically decays as $q/r^{2}+G\ \mathcal{O}(r^{-4})$. Note that in an
orthonormal frame the electric field, however, has the usual form, $%
F^{01}=q/r^{2}$.

The scalar field at the horizon is given by 
\begin{equation*}
\left. \phi \right\vert _{r=r_{+}}=\sqrt{\frac{3}{4\pi G}}\;\mbox{Arctanh}%
\left( \sqrt{G}\frac{|q|}{l}\right) \ ,
\end{equation*}%
and asymptotically behaves as%
\begin{equation}
\phi =\sqrt{\frac{3}{4\pi }}\frac{|q|}{r}+\mathcal{O}(r^{-3})\ .
\label{Asympt-Phi}
\end{equation}%
As discussed in \cite{HMTZ-2+1}, and further developed in \cite%
{HMTZ-Log,Hertog-Maeda,HMTZ-D}, the presence of scalar fields with a slow
fall off as in Eq. (\ref{Asympt-Phi}) has generically two effects: It gives
rise to a strong back reaction that relaxes the standard asymptotic form of
the geometry, and it generates additional contributions to the charges that
depend explicitly on the scalar fields at infinity which are not already
present in the gravitational part. These effects has also been discussed
recently following the covariant phase space method \cite{Marolf}. The
asymptotic form of the metric (\ref{Black-Hole}) reads%
\begin{eqnarray*}
g_{tt} &=&-\left( \frac{r^{2}}{l^{2}}-1\right) +\mathcal{O}(r^{-2}) \\
g_{rr} &=&\frac{l^{2}}{r^{2}}+\left( 1-3G\frac{q^{2}}{l^{2}}\right) \frac{%
l^{4}}{r^{4}}+\mathcal{O}(r^{-6})
\end{eqnarray*}%
which manifestly deviates from the standard behavior \cite%
{Henneaux-Teitelboim}.

The mass of the black hole under consideration can be computed explicitly
from a surface integral as in \cite{HMTZ-D}. For a scalar field satisfying $%
m^{2}=\left. V^{\prime \prime }\right\vert _{\phi =0}=-2l^{-2}$, the mass is
given by%
\begin{equation}
M=Q_{G}(\partial _{t})+Q_{\phi }(\partial _{t})\ ,  \label{Mass}
\end{equation}%
where $Q_{G}(\partial _{t})$ stands for the standard formula \cite%
{Abbott-Deser,Henneaux-Teitelboim}, and the contribution from the scalar
field is 
\begin{equation}
Q_{\phi }(\partial _{t})=\frac{1}{6}\int d^{2}\sigma r^{3}\left[ \left( 
\frac{r}{l}\partial _{r}\phi \right) ^{2}-m^{2}\phi ^{2}-\frac{1}{3}\left.
V^{\prime \prime \prime }\right\vert _{\phi =0}\ \phi ^{3}\right] \ .
\label{Q-Phi}
\end{equation}%
For the potential (\ref{Potential}), the coefficient $\left. V^{\prime
\prime \prime }\right\vert _{\phi =0}$ of the cubic term in (\ref{Q-Phi})
vanishes, and as expected, evaluating (\ref{Mass}) for the black hole, the
divergences coming from $Q_{G}$ and $Q_{\phi }$ are cancelled. Thus the
mass, which is given by the remaining finite term, is found to vanish. This
can also be seen from the asymptotic form of the fields, since the terms
coming from the metric that contribute to mass, which are the ones of order $%
r^{-1}$ in $g_{tt}$, and order $r^{-5}$ in $g_{rr}$, are absent. Moreover,
the contribution to the mass coming from the scalar field requires that both
leading orders in the scalar field, i.e., the orders $r^{-1}$ and $r^{-2}$,
must be simultaneously present. Therefore, the contribution to the mass
coming from the scalar field also vanishes since the term of order $r^{-2}$
does not appear in the asymptotic form of $\phi $ given by Eq. (\ref%
{Asympt-Phi}). These results could also be discussed following different
covariant approaches as in \cite{Glenn} and \cite{Chen-Lu-Pope}.

\section{Thermodynamics}

\label{thermo}

The thermodynamics for the electrically charged black hole with scalar hair
is discussed using the Euclidean approach. For this purpose it is useful to
consider a minisuperspace of static Euclidean metrics given by%
\begin{equation}
ds^{2}=N(r)^{2}f(r)^{2}dt^{2}+f(r)^{-2}dr^{2}+r^{2}d\sigma^{2},  \label{dse}
\end{equation}%
where the Euclidean time has period $\beta $, and the radius ranges as $%
r\geq r_{+}$. The scalar and electromagnetic fields are assumed to be of the
form $\phi =\phi (r)$, and $A=A_{t}(r)dt$, respectively. The temperature $T$
corresponds to the inverse of $\beta $, and it is fixed requiring that the
allowed class of the geometries (\ref{dse}) should contain no conical
singularities at the horizon. This condition implies that 
\begin{equation}
\left. \beta (N(r)(f^{2}(r))^{\prime })\right\vert _{r=r_{+}}=4\pi \;,
\label{regular}
\end{equation}%
which for the black hole solution (\ref{Black-Hole}), yields the following
temperature 
\begin{equation}
T=\beta ^{-1}=\frac{1}{2\pi l}\;.  \label{beta}
\end{equation}%
Note that the temperature is determined only by the AdS radius and it is
independent of the horizon size.

The Euclidean path integral in the saddle point approximation around the
Euclidean solution is identified with the partition function of a
thermodynamical ensemble \cite{Gibbons:1976ue}. Here we consider the
Euclidean continuation of the action (\ref{action}) in Hamiltonian form.
Following the analysis performed in Ref. \cite{MTZ-Top}, and including
electric charge, one obtains that the reduced Hamiltonian action is 
\begin{equation}
I=-\frac{\beta \sigma }{4\pi }\int_{r_{+}}^{\infty }\left( N(r)\mathcal{H}%
(r)+A_{t}p^{\prime }\right) dr+B\ ,  \label{rhaction}
\end{equation}%
where $B$ is a surface term, and $\sigma $ is the area of the base manifold $%
\Sigma $. The reduced Hamiltonian is given by%
\begin{equation*}
\mathcal{H}=\frac{r^{2}}{2G}\left[ \left( \frac{(f^{2})^{\prime }}{r}+\frac{1%
}{r^{2}}(1+f^{2})\right) +4\pi G\left( f^{2}(\phi ^{\prime })^{2}+2V(\phi
)\right) +\frac{Gp^{2}}{r^{4}}\right] \ ,
\end{equation*}%
and $p$ is defined in terms electric field as%
\begin{equation*}
p=\frac{r^{2}}{N}A_{t}^{\prime }\ .
\end{equation*}%
The Euclidean black hole solution is static and satisfies the constraints $%
\mathcal{H}=0$, $p^{\prime }=0$. Therefore, the action (\ref{rhaction})
evaluated on the classical solution is just given by the boundary term $B$.
The boundary term is fixed by requiring the action (\ref{rhaction}) to
attain an extremum for the minisuperspace considered here \cite%
{Regge-Teitelboim}.

In what follows, we work in the grand canonical ensemble, so that we
consider variations of the action keeping fixed the temperature and the
\textquotedblleft voltage", i.e., $\beta $ and $\Phi =A_{t}(\infty
)-A_{t}(r_{+})$ are constants.

The variation of the required boundary term is 
\begin{equation}
\delta B\equiv \delta B_{G}+\delta B_{\phi }+\delta B_{em}\;,
\label{delBtotal}
\end{equation}%
where 
\begin{equation}
\delta B_{G}=\frac{\beta \sigma }{8\pi G}\left[ Nr\delta f^{2}\right]
_{r_{+}}^{\infty }\;,  \label{delG}
\end{equation}%
and the contribution from the matter sector is given by 
\begin{eqnarray}
\delta B_{\phi } &=&\beta \sigma Nr^{2}f^{2}\phi ^{\prime }\delta \phi
|_{r_{+}}^{\infty }\;.  \label{delphi} \\
\delta B_{em} &=&\frac{\beta \sigma }{4\pi }\Phi \delta p\ .  \label{delem}
\end{eqnarray}%
The variation of the fields at infinity for the black hole solution (\ref%
{Black-Hole}, \ref{scalar}, \ref{A}) reads 
\begin{eqnarray}
\left. \delta f^{2}\right\vert _{\infty } &=&\frac{3G}{l^{2}}\delta
q^{2}+O\left( r^{-2}\right) \;, \\
\left. \delta \phi \right\vert _{\infty } &=&\sqrt{\frac{3}{\pi }}\frac{1}{%
4|q|r}\delta q^{2}+O\left( r^{-3}\right) \;, \\
\left. \delta p\right\vert _{\infty } &=&\delta q
\end{eqnarray}%
and thus, one obtains 
\begin{equation}
\left. \delta B_{G}\right\vert _{\infty }=\frac{3\beta \sigma }{8\pi l^{2}}\
r\delta q^{2}+O\left( r^{-1}\right) \;.  \label{B2}
\end{equation}%
The variation of the purely gravitational contribution to the boundary term $%
\left. \delta B_{G}\right\vert _{\infty }$ has linearly divergent term and
is devoid of a finite piece. As discussed above, this reflects the fact that
the scalar field produces a slow decay for the metric as compared with that
of pure gravity with a standard localized distribution of matter \cite%
{Henneaux-Teitelboim}. This divergence is cancelled by the contribution
coming from the scalar field 
\begin{equation}
\left. \delta B_{\phi }\right\vert _{\infty }=-\frac{3\beta \sigma }{8\pi
l^{2}}\ r\delta q^{2}+O\left( r^{-1}\right) \;.
\end{equation}%
Choosing $\left. A\right\vert _{\infty }$ to vanish, the total boundary term
at infinity then vanishes%
\begin{equation}
\left. B\right\vert _{\infty }=0\ .  \label{Binf}
\end{equation}%
The variation of the boundary term at the horizon, is obtained using 
\begin{equation*}
\left. \delta f^{2}\right\vert _{r_{+}}=-\left. (f^{2})^{\prime }\right\vert
_{r_{+}}\delta r_{+}\;,
\end{equation*}%
and Eqs. (\ref{regular}, \ref{delG}, \ref{delphi}, \ref{delem}). Since, $%
\left. \delta B_{\phi }\right\vert _{r_{+}}$ vanishes, the variation of the
total boundary term is 
\begin{eqnarray}
\left. \delta B\right\vert _{r_{+}} &=&\frac{\sigma }{4\pi }\left( -\frac{1}{%
4G}N(r_{+})\beta \left. (f^{2})^{\prime }\right\vert _{r_{+}}\delta \left(
r_{+}^{2}\right) +\beta \Phi \delta p\right)  \label{delBtotal2} \\
&=&-\frac{\sigma }{4G}\delta \left( r_{+}^{2}\right) -\frac{\beta \sigma }{%
4\pi }\Phi \delta q\ ,  \notag
\end{eqnarray}%
where the last term in (\ref{delBtotal2}) is the contribution from the
electric field. Hence, the boundary term at the horizon can be integrated as%
\begin{equation}
\left. B\right\vert _{r_{+}}=-\frac{\sigma }{4G}r_{+}^{2}-\frac{\beta \sigma 
}{4\pi }\Phi q\ .  \label{Bhor}
\end{equation}

The value of the Euclidean action on shell is then given by the boundary
terms in Eqs. (\ref{Binf}) and (\ref{Bhor}), which reads 
\begin{equation}
I=\frac{\sigma }{4G}r_{+}^{2}+\frac{\beta \sigma }{4\pi }\Phi q\;,
\label{Ieval}
\end{equation}%
up to an arbitrary additive constant. The Euclidean action is related to the
free energy (in units where $\hbar =k_{B}=1$) as $I=-\beta F$, which in the
grand canonical ensemble is given by%
\begin{equation}
I=S-\beta M+\beta \Phi Q\;.  \label{appro}
\end{equation}%
Here $M$, $Q$ and $S$ stand for the mass, electric charge and entropy,
respectively. Thus, once the free energy is identified with the Euclidean
action, these quantities must satisfy the first law of thermodynamics.
Expressions (\ref{Ieval}) and (\ref{appro}), allow to obtain the mass the
electric charge and the entropy from%
\begin{eqnarray*}
M &=&\left( \beta ^{-1}\Phi \frac{\partial }{\partial \Phi }-\frac{\partial 
}{\partial \beta }\right) I=0\ , \\
Q &=&\beta ^{-1}\frac{\partial I}{\partial \Phi }=\frac{\sigma }{4\pi }q\ ,
\\
S &=&\left( 1-\beta \frac{\partial }{\partial \beta }\right) I=\frac{\sigma 
}{4G}r_{+}^{2}\ .
\end{eqnarray*}%
Thus, as expected, the entropy follows the area law since the horizon area
is given by $\sigma r_{+}^{2}$, and the mass is shown to vanish from an
independent method.

\section{Thermal decay}

\label{thermal}

For a fixed temperature and electromagnetic potential, the action principle (%
\ref{action}), also admits an electrically charged solution for the same
boundary conditions but without hair, i.e. with $\phi \equiv 0$. This
solution \cite{Brill-Louko-Peldan} is described by 
\begin{equation}
ds^{2}=-\left[ \frac{\rho ^{2}}{l^{2}}-1-\frac{2G\mu _{0}}{\rho }+\frac{%
Gq_{0}^{2}}{\rho ^{2}}\right] dt^{2}+\left[ \frac{\rho ^{2}}{l^{2}}-1-\frac{%
2G\mu _{0}}{\rho }+\frac{Gq_{0}^{2}}{\rho ^{2}}\right] ^{-1}d\rho ^{2}+\rho
^{2}d\sigma ^{2}\;,  \label{phizero}
\end{equation}%
with 
\begin{equation}
A=-\frac{q_{0}}{\rho }dt\ ,  \label{A0}
\end{equation}%
for which the mass and the electric charge by $M_{0}=\sigma \mu _{0}/4\pi $,
and $Q_{0}=\sigma q_{0}/4\pi $.

Since the hairy black hole (\ref{Black-Hole}) has a fixed temperature given
by (\ref{beta}), the matching of this temperature with the one for the black
hole in Eq. (\ref{phizero}) reads\footnote{%
Note that, since the base manifold is locally hyperbolic, at fixed
temperature and voltage, there is only one black hole without scalar hair.
For spherical symmetry there are two possible black holes without scalar
hair satisfying these requirements.}%
\begin{equation*}
\beta _{0}=\frac{4\pi \rho _{+}}{3\frac{\rho _{+}^{2}}{l^{2}}-\frac{%
q_{0}^{2}G}{\rho _{+}^{2}}-1}=\beta =2\pi l\ ,
\end{equation*}%
where $\beta _{0}$ stands for the Euclidean period of the black hole without
scalar hair. This condition is fulfilled when the mass and the electric
charge of the black hole without scalar hair relate with the horizon radius $%
\rho _{+}$, in the following way: 
\begin{eqnarray}
G\mu _{0}l &=&\rho _{+}^{2}\left( \frac{2\rho _{+}}{l}-\frac{l}{\rho _{+}}%
-1\right) \ ,  \notag \\
Gq_{0}^{2} &=&\frac{\rho _{+}^{3}}{l}\left( \frac{3\rho _{+}}{l}-\frac{l}{%
\rho _{+}}-2\right) \ .  \label{Match1}
\end{eqnarray}%
Analogously, matching the voltages amounts to match the electromagnetic
potentials (\ref{A}) and (\ref{A0}) at the horizon, which leads to

\begin{equation}
\frac{q}{l}=\frac{q_{0}}{\rho _{+}}\ .  \label{Match2}
\end{equation}%
Note that since $q_{0}^{2}\geq 0$, both black holes can have the same
temperature\footnote{%
This bound is saturated for $\mu _{0}=q_{0}=0$, but in this case, the
matching of the voltages implies that $q=0$ and $r_{+}=l$. This means that
when the bound is saturated, matter fields are switched off and both metrics
coincide.} provided $\rho _{+}\geq l$. This raises the question of whether
one black hole can decay into the other. Since the partition function is
given by $Z=\exp (I)$, this can be examined evaluating the difference
between the corresponding Euclidean actions.

The Euclidean action evaluated on the hairy black hole (\ref{Black-Hole}, %
\ref{A}, \ref{scalar}) reads%
\begin{equation}
I_{\phi }=\frac{\sigma l^{2}}{4G}\left[ 1+G\frac{q^{2}}{l^{2}}\right] \ ,
\label{IPhi}
\end{equation}%
which by virtue of the matching conditions (\ref{Match1}) and (\ref{Match2}%
), can be expressed as%
\begin{equation*}
I_{\phi }=\frac{\sigma }{2G}\rho _{+}^{2}\left[ \frac{3}{2}-\frac{l}{\rho
_{+}}\right] \ .
\end{equation*}%
For the black hole without scalar hair the Euclidean action is given by 
\begin{equation*}
I_{0}=\frac{\sigma }{2G}\rho _{+}^{2}\left[ \frac{\rho _{+}}{l}-\frac{1}{2}%
\right] \ .
\end{equation*}%
Therefore, the difference between both Euclidean actions is%
\begin{equation}
\Delta I=I_{0}-I_{\phi }=\frac{\sigma l}{2G}\rho _{+}\left[ \frac{\rho _{+}}{%
l}-1\right] ^{2}\ ,  \label{DeltaI}
\end{equation}%
which is always positive. This means that there is a nonvanishing
probability for the decay of the hairy black hole into the black hole
without scalar field, so that the black hole without scalar hair is
thermodynamically favored.

For a fixed temperature and voltage, by virtue of (\ref{Match1}), the
difference between the black hole masses under the decay is always positive
for the allowed range, $\rho _{+}>l$. Analogously, the difference between
the absolute values of the electric charges is%
\begin{equation*}
\Delta |q|=|q_{0}|-|q|=|q_{0}|\left( 1-\frac{l}{\rho _{+}}\right) >0\ .
\end{equation*}%
Similarly, since the entropy for the black hole without scalar field is $%
S_{0}=\sigma \rho _{+}^{2}(4G)^{-1}$, the entropies for the allowed range
are found to obey 
\begin{equation*}
\Delta S=S_{0}-S_{\phi }=\frac{\sigma l^{2}}{2}\frac{\mu _{0}}{\rho _{+}}>0\
.
\end{equation*}%
In sum, as $\Delta I$ in Eq. (\ref{DeltaI}) is positive, there is a
nonvanishing probability for the decay of the black hole dressed with the
scalar field into the bare black hole. Since the process take place for
black holes in vacuum with $\rho _{+}>l$, which has positive mass, in the
decay process, the scalar black hole absorbs energy and electric charge from
the thermal bath, increasing its horizon radius an consequently its entropy.
This suggests that in this process the scalar field is at least partially,
absorbed by the black hole.

\subsection{Transitions involving the extremal charged black hole without
hair}

The black hole without scalar hair described by Eqs. (\ref{phizero}), (\ref%
{A0}) admits an extremal case, for which the mass and the electric charge
are fine tuned such that its temperature vanishes, and thus the Euclidean
time period $\beta _{e}$ is arbitrary. This opens two additional possible
decay channels to be explored.

\subsubsection{Stability of the non-extremal charged black hole without
scalar hair}

Let us begin analyzing the transition between the extreme and non-extreme
charged black holes without scalar hair. Following the same procedure as in
Section \ref{thermo}, but taking into account that the Euclidean period is
arbitrary, it is found that the entropy vanishes, as in Ref. \cite%
{Teitelboim-Entropy} (see also \cite{Gibbons-Kallosh,Hawking-Horowitz-Ross}%
), and the Euclidean action is given by

\begin{equation}
I_{e}=\beta _{e}\frac{\sigma }{4\pi Gl^{2}}\rho _{e}^{3}\ ,  \label{Ie}
\end{equation}%
where $\rho _{e}$ is the horizon radius of the extremal black hole.

For the non extremal black hole without scalar field the Euclidean action
can be expressed $q_{0}$ and $\rho _{+}$ as%
\begin{equation*}
I_{0}=\frac{\sigma \rho _{+}^{2}}{4G}\left[ \frac{1+\frac{\rho _{+}^{2}}{%
l^{2}}+\frac{q_{0}^{2}G}{\rho _{+}^{2}}}{3\frac{\rho _{+}^{2}}{l^{2}}-\frac{%
q_{0}^{2}G}{\rho _{+}^{2}}-1}\right] \ .
\end{equation*}%
For the decay process, the Euclidean time period of the extremal black hole
is then fixed as $\beta _{e}=\beta _{0}$, and matching the voltages leads to%
\begin{equation*}
\frac{q_{0}}{\rho _{+}}=\frac{q_{e}}{\rho _{e}}\ .
\end{equation*}%
This allows to express the Euclidean action for the extremal black hole (\ref%
{Ie}) in terms of the horizon radius and the electric charge of the
non-extremal black hole as%
\begin{equation*}
I_{e}=\frac{\sigma l}{G}\rho _{+}\left[ \frac{\left( \frac{1}{3}\left[ 1+%
\frac{Gq_{0}^{2}}{\rho _{+}^{2}}\right] \right) ^{\frac{3}{2}}}{3\frac{\rho
_{+}^{2}}{l^{2}}-\frac{q_{0}^{2}G}{\rho _{+}^{2}}-1}\right] \ .
\end{equation*}%
Thus, the difference between both Euclidean actions is%
\begin{equation*}
\Delta I=I_{0}-I_{e}=\frac{\sigma }{4G}\frac{\rho _{+}^{2}+\frac{\rho
_{+}^{4}}{l^{2}}+q_{0}^{2}G-4l\rho _{+}\left( \frac{1}{3}\left[ 1+\frac{%
Gq_{0}^{2}}{\rho _{+}^{2}}\right] \right) ^{\frac{3}{2}}}{3\frac{\rho
_{+}^{2}}{l^{2}}-\frac{q_{0}^{2}G}{\rho _{+}^{2}}-1}\ ,
\end{equation*}%
which can be shown to be positive for the allowed range of the parameters.
This ensures the stability of the non-extreme black hole, and one then
concludes that there is a non vanishing probability for the decay of the
extremal into the non-extremal solution. This result is qualitatively
similar with what was found in Ref. \cite{Cai-hyperbolic}.

\subsubsection{Spontaneous scalar field dressing up of the extremal black
hole}

Consider now the transition between the hairy and the extreme case. The
Euclidean action for the hairy black hole $I_{\phi }$ is given by Eq. (\ref%
{IPhi}), and the one for the extremal solution without scalar field $I_{e}$
is given by (\ref{Ie}). In this case, according to Eq. (\ref{beta}), the
Euclidean period of the extremal solution must be fixed as $\beta _{e}=\beta
_{\phi }=2\pi l$, and the matching the voltages reads%
\begin{equation}
\frac{q}{r_{+}}=\frac{q_{e}}{\rho _{e}}\ .  \label{LastMatch}
\end{equation}%
Since the Euclidean action $I_{\phi }$ in (\ref{IPhi}) depends only on the
electric charge it is convenient to make use of (\ref{horizon}) and (\ref%
{LastMatch}) in order to express the Euclidean action $I_{e}$ in (\ref{Ie})
as%
\begin{equation*}
I_{e}=\frac{\sigma l^{2}}{2G}\left[ \frac{1}{3}\left( \frac{l^{2}}{%
l^{2}-Gq^{2}}\right) \right] ^{\frac{3}{2}}\ .
\end{equation*}%
Therefore, the difference between the Euclidean action of the hairy black
hole and the one for the extremal case without scalar hair is given by%
\begin{equation}
\Delta I=I_{\phi }-I_{e}=\frac{\sigma l^{2}}{2G}\left( \frac{1}{2}+\frac{%
Gq^{2}}{2l^{2}}-\left[ \frac{1}{3}\left( \frac{l^{2}}{l^{2}-Gq^{2}}\right) %
\right] ^{\frac{3}{2}}\right) \ ,  \label{DeltaILast}
\end{equation}%
which changes of sign for a critical value of the electric charge $q_{c}$
satisfying%
\begin{equation*}
\frac{Gq_{c}^{2}}{l^{2}}\approx 0.615713\ .
\end{equation*}%
Remarkably, for $q^{2}<q_{c}^{2}$, the difference of Euclidean actions (\ref%
{DeltaILast}) is positive, and hence the extremal bare black hole likely to
undergo a spontaneous dressing up with a nontrivial scalar field. For $%
q^{2}>q_{c}^{2}$, it turns out that $\Delta I<0$, and then there is a non
vanishing probability for the decay of hairy black hole decay into the
extremal bare solution. For the critical point, $q^{2}=q_{c}^{2}$, both
solutions can coexist.

\section{Concluding remarks}

\label{concluding}

It was shown that gravity minimally coupled to a real self-interacting
scalar field and electromagnetism in four dimensions admits a charged hairy
black hole solution. The event horizon is a surface of negative constant
curvature and the asymptotic region has negative constant curvature. The
self-interacting potential (\ref{Potential}) is negative and unbounded from
below, possessing global maximum at $\phi =0$, and it has a mass term
satisfying the Breitenlohner-Freedman bound that guarantees the perturbative
stability of global AdS spacetime \cite{B-F}. For the topology considered
here, it was shown that the stability of the locally AdS spacetime (\ref%
{muzero}) under scalar perturbations, holds provided the mass satisfies the
same BF bound \cite{Aros:2002te}. The asymptotic fall-off of the fields is
slower than the standard one, and for the scalar field, only the branch with
the slower fall-off is switched on. In spite of this, the mass is still well
defined and it is shown to vanish. The scalar field is regular everywhere
except at the origin, and is supported by the presence of electric charge
which is bounded from above by the AdS radius. An upper bound for the
electric charge also exists for the massless bare black hole with the same
topology.

The presence of the real scalar field smooths the electromagnetic potential
everywhere. The hairy black hole has a fixed temperature $T=(2\pi l)^{-1}$
regardless the value of the electric charge, and the entropy follows the
usual area law. It is also shown that there is a nonvanishing probability
for the decay of the hairy black hole into a black hole without scalar
field, where in the decay process, the hairy black hole absorbs energy and
electric charge from the thermal bath, increasing its horizon radius an
consequently its entropy. A similar behavior has been previously observed
for black holes with scalar hair in Refs. \cite{GMT,Hertog-MaedaSTAB}. It is
worth pointing out that, although the boundary conditions considered here
coincide with the ones for the hairy black hole found in \cite{MTZ-Top},
there is no phase transition where the bare non-extremal black hole
spontaneously dresses up with a scalar field. Furthermore, since the bare
black hole admits an extremal case, for which the Euclidean time period is
arbitrary, two additional decay channels open up. Noteworthy, it is found
that, an extremal black hole without scalar field is likely to undergo a
spontaneous dressing up with a nontrivial scalar field, provided the
electric charge satisfies $q^{2}<q_{c}^{2}$; while for $q^{2}>q_{c}^{2}$,
there is a non vanishing probability for the decay of hairy black hole decay
into the extremal bare solution.

In sum, for a fixed voltage and temperature $\beta =2\pi l$, one obtains
that if the electric charge of the hairy black hole satisfies $%
q^{2}<q_{c}^{2}$, then the Euclidean actions satisfy:

\begin{equation*}
I_{0}>I_{\phi }>I_{e}\ ,
\end{equation*}%
which means that the non-extremal bare black hole is stable. Consequently,
the hairy black hole is likely to decay into non-extremal one, and moreover,
the extremal black hole is able to decay into the hairy or into the
non-extremal black hole, with different branching ratios. For $%
q^{2}>q_{c}^{2}$, the Euclidean actions fulfill:

\begin{equation*}
I_{0}>I_{e}>I_{\phi }\ ,
\end{equation*}%
which means that for this range, the scalar black hole can decay into the
extreme or into the non-extreme black hole with different probabilities. In
the case $q^{2}=q_{c}^{2}$, both the hairy and the extremal black hole can
coexist, but they can decay into the non-extremal solution without scalar
field.

As a final remark, it is worth pointing out that in the conformal frame, the
potential (\ref{Potential}) is mapped to a negative cosmological constant,
and the scalar field becomes conformally coupled without self interaction.
Following \cite{MST}, it can also be seen that in the conformal frame, the
black hole solves the vacuum field equations since the stress-energy tensor
for the scalar field cancels with the one for the electromagnetic field. The
effect of having nontrivial matter fields with a vanishing total energy
momentum tensor have also been discussed for flat spacetime in \cite%
{Cheshire,Robinson} and for three-dimensional gravity with negative
cosmological constant in Refs. \cite%
{Natsuume+Eloy+HMTZ2+1+Jack+Stealth,HMTZ-2+1,GMT}. This effect has also been
discussed for different setups in Refs. \cite{Mokhtar+Olivera+Demir}.

\textbf{Acknowledgments}

We thank J. Zanelli for useful discussions and comments. This research is
partially funded by FONDECYT grants N$^o$ 1051064, 1040921, 1051056,
1061291. The generous support to Centro de Estudios Cient\'{\i}ficos (CECS)
by Empresas CMPC is also acknowledged. CECS is funded in part by grants from
the Millennium Science Initiative, Fundaci\'{o}n Andes and the Tinker
Foundation.

\end{document}